\documentclass[twocolumn,preprintnumbers,amsmath,amssymb]{revtex4}
\usepackage{epsfig}
\begin{document}
\title{Diffusion of Wilson Loops}
\preprint{FAU-TP3-04/7}
\preprint{MIT-CTP-3569}
\author{A. M. Brzoska$^{{\rm a}}$}
\author{F. Lenz$^{{\rm a}}$}
\author{J. W. Negele$^{{\rm b}}$}
\author{M. Thies$^{{\rm a}}$}
\affiliation{$^{{\rm a}}$ Institut f\"ur Theoretische Physik III,
Universit\"at Erlangen-N\"urnberg, D-91058 Erlangen, Germany \\
$^{{\rm b}}$ Center for Theoretical Physics, Laboratory for Nuclear Science,\\
and Department of Physics, Massachusetts Institute of Technology, Cambridge,
Massachusetts 02139, USA}
\date{\today}
\begin{abstract} A phenomenological analysis of the distribution of Wilson loops in SU(2) Yang-Mills theory
is presented  in which Wilson loop distributions are described as the  result of a diffusion process  on the
group manifold. It is shown that, in the absence of  forces,  diffusion implies Casimir scaling and, conversely,
exact Casimir scaling implies free diffusion. Screening processes occur if diffusion takes place in a potential.
The crucial distinction  between  screening of fundamental and adjoint loops is formulated as a symmetry
property related to the center symmetry of the underlying gauge theory. The results are expressed in terms
of  an effective Wilson loop action  and compared with various limits of SU(2) Yang-Mills theory. 
\end{abstract}
\maketitle
\section{Introduction}
Wilson loops play an important role in gauge theories. They describe closed gauge strings. Their expectation
value characterizes the phases of Yang-Mills theories. Extensive  calculations of Wilson loops in lattice gauge
 theories have demonstrated confinement and shown the detailed behavior of the static quark potential. They
 have also revealed the existence of an intermediate dynamical regime where simultaneously confinement
  prevails and so-called ``Casimir scaling'' occurs, in which the string tension in each representation is proportional to
the quadratic Casimir operator  \cite{amop84}. Asymptotically, for large loops,  Casimir scaling must break down
 and be replaced by  complete screening of adjoint charges and screening of half-integer charges to fundamental 
charges. In lattice calculations for SU(3) Yang-Mills theory, significant deviation from Casimir scaling has not yet
 been observed \cite{bali00} while in SU(2), indications of  a transition to  the screening regime have been obtained 
\cite{defo00,katr02}. 

Although these  results of lattice gauge calculations have not yet produced a breakthrough
 in the understanding of the dynamics of Wilson loops,  they have raised new questions. In particular,  the origin
 of Casimir scaling turns out to be quite mysterious. Casimir scaling is expected theoretically for loops
 small enough to be calculable in lowest order perturbation theory, or may occur in related approximation schemes
such as the lowest order term of the cumulant expansion of the Wilson loop expectation value \cite{Shoshi02}.
However a generally accepted explanation for the observed Casimir
 scaling in the confinement regime does not exist. Two-dimensional Yang-Mills theories are known to exhibit Casimir scaling for
 loops of arbitrary size and one may invoke ``dimensional reduction'' \cite{amop84} to relate Yang-Mills theories in
 two and four dimensions. However, in two dimensions  no  screening occurs, and therefore the limit of asymptotically
 large loops is fundamentally  different from the four-dimensional case. Wilson loops are  complicated non-local, 
composite objects, which makes  understanding their  properties in terms of the fundamental degrees of freedom
 difficult.  Hence, it is useful to take a phenomenological approach. Based on the observation that Wilson loop 
distributions calculated in lattice QCD coincide, to a high degree of accuracy, with distributions resulting from a 
diffusion process, we introduce a phenomenology based on diffusion on the group manifold.  The resulting theoretical 
description  contains Casimir scaling as a natural limit and leads to a natural generalization that incorporates 
screening-induced deviations from this limit. This approach leads to the formulation of an effective action for
 Wilson loops.
\section{Wilson loops}
\begin{figure*}
\hfill\epsfig{file=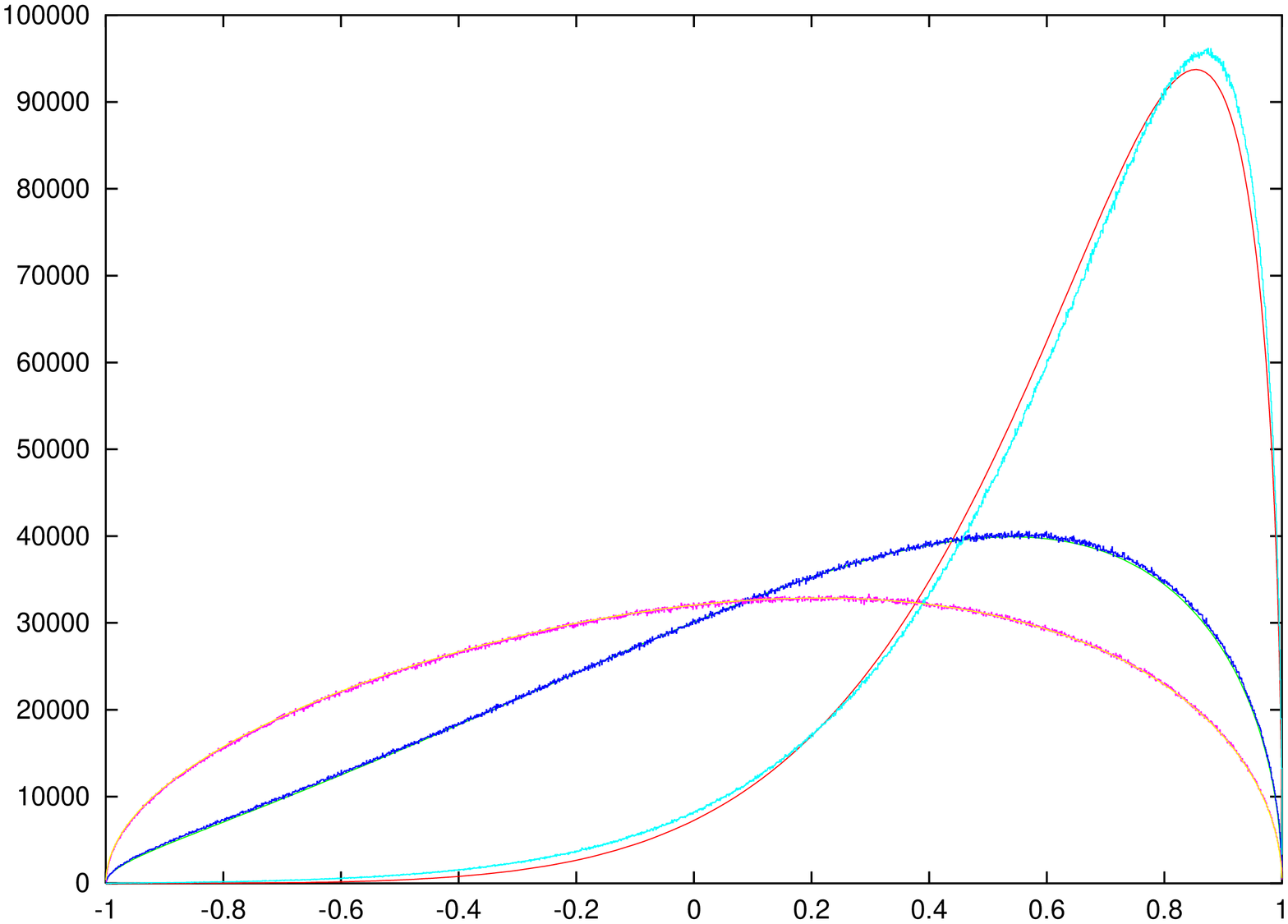, width=0.3\linewidth, angle=-90}\hfill\hfill\epsfig{file=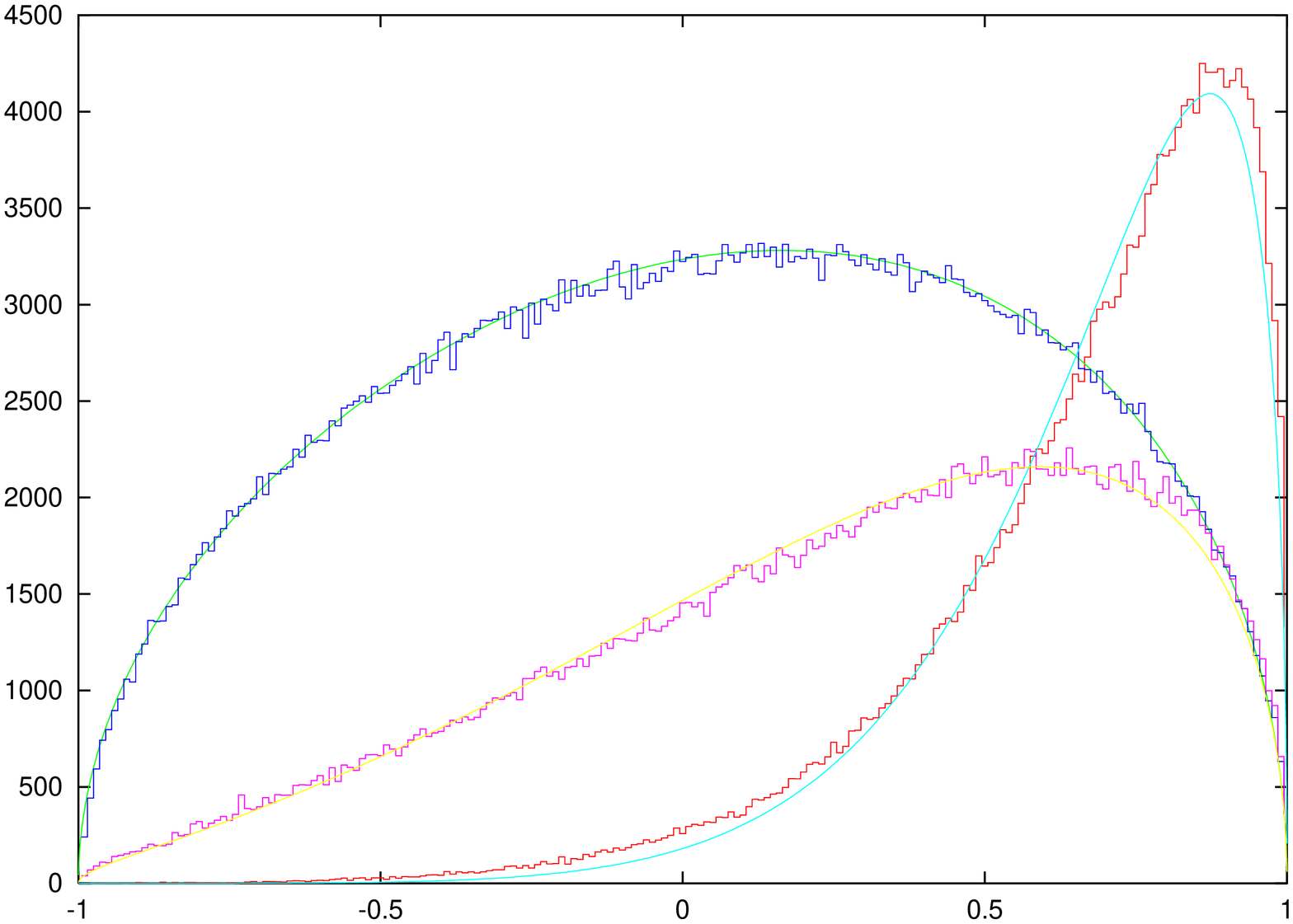, width=.3\linewidth,angle=-90}\hfill\
\caption{Wilson loop distribution for rectangular loops with aspect ratio 2 of different areas in SU(2) lattice gauge
theory (left) and for an ensemble of field configurations containing 500 merons (right). These results have been fitted
 with distributions resulting from diffusion, Eq.~(\ref{gp}), denoted by dashed lines.}
\label{wdlm}
\end{figure*}
The object of our investigations is the distribution of Wilson loops of fixed geometry,
\begin{equation}
\label{siwi}
p(\omega) = \langle 0|\delta \left(\omega - W\right) |0\rangle \ ,
\end{equation}
in SU(2) Yang-Mills theory. Wilson loops  describe  closed gauge strings  along a prescribed  curve ${\cal C}$
\begin{equation}
\label{wilo1}
 W =  \frac{1}{2} \mbox{tr}\, P\, \exp\{{\rm i}g\oint_{{\cal C}} {\rm d}x^{\mu}A_{\mu}(x)\}\, .
\end{equation}
In the confining phase, the vacuum expectation values of Wilson loops of sufficiently large area, ${\cal A}$, obey
the area law
\begin{equation}
\label{area}
\langle 0|W|0\rangle \sim {\rm e}^{-\sigma {\cal A}}\ ,    
\end{equation}
where  $\sigma$ denotes the  string tension. The expectation value of  a rectangular loop with side $T$ much
larger than side $R$ is   determined by  the interaction energy  of static quarks, $V(R)$
\begin{equation}
\label{statq}
\langle 0| W|0\rangle \sim {\rm e}^{-T\, V(R)}\ .  
\end{equation}
As illustrated  in Fig.~\ref{wdlm}, for small sizes, the Wilson loop distribution is peaked at values close to 1,
 whereas, for large sizes,  it  approaches  the Haar measure
\begin{equation}
{\cal A} \to \infty\, , \quad p(\omega)\to \frac{2}{\pi}\sqrt{\,1-\omega^2\,}\, .
\label{hm}
\end{equation}
\section{Diffusion of  Wilson Loops}
Our description of the Wilson loop distribution starts with the observation that for a fixed gauge field 
configuration, either from lattice QCD or an ensemble of merons or regular gauge instantons,  the value
 of a Wilson loop of a given shape and orientation undergoes a Brownian-like random walk as the size of the
 loop is progressively increased. 

\begin{figure*}[t]\hfill\epsfig{file=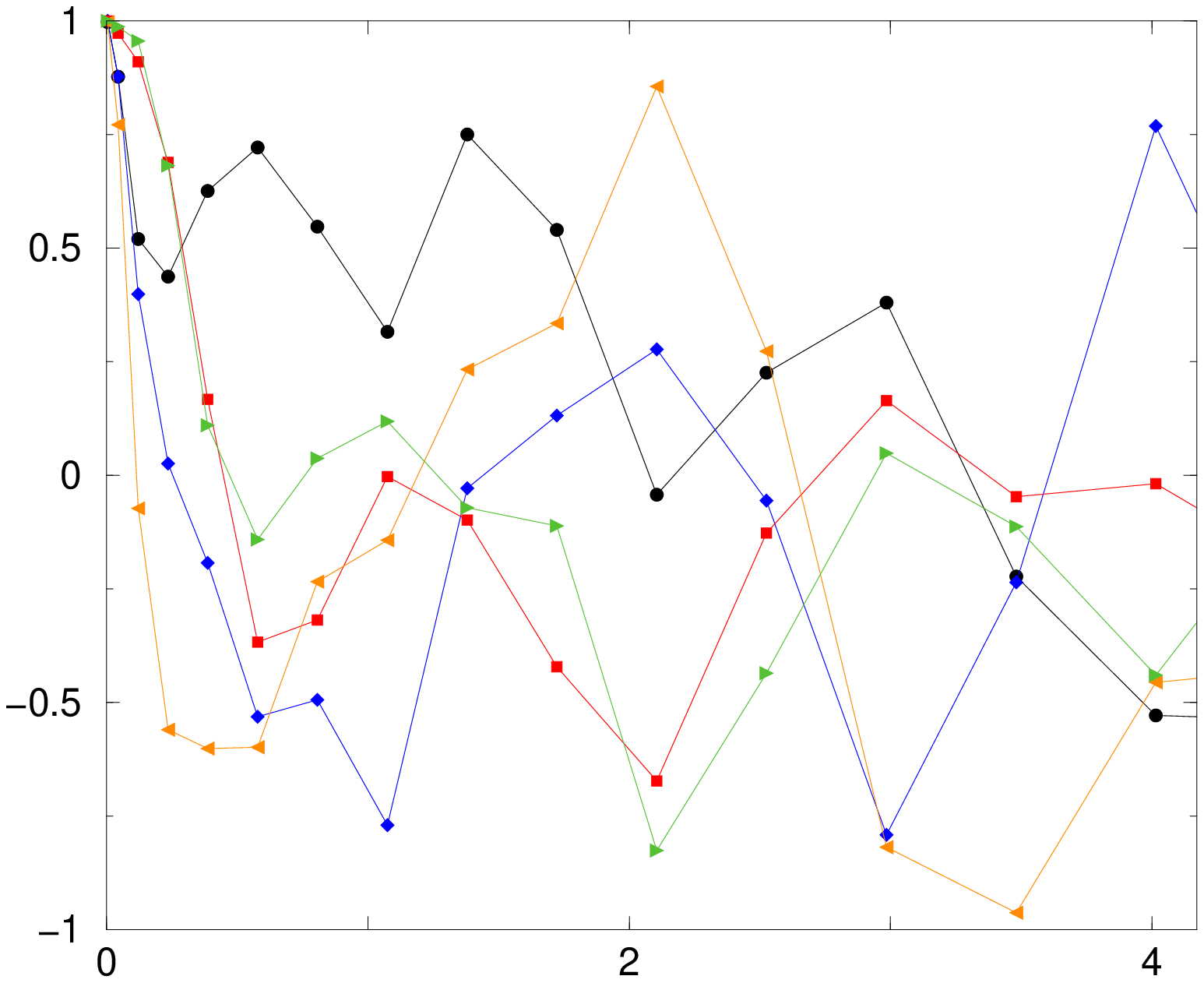, width=0.4\linewidth,height=0.285\linewidth}
\hfill\hfill\epsfig{file=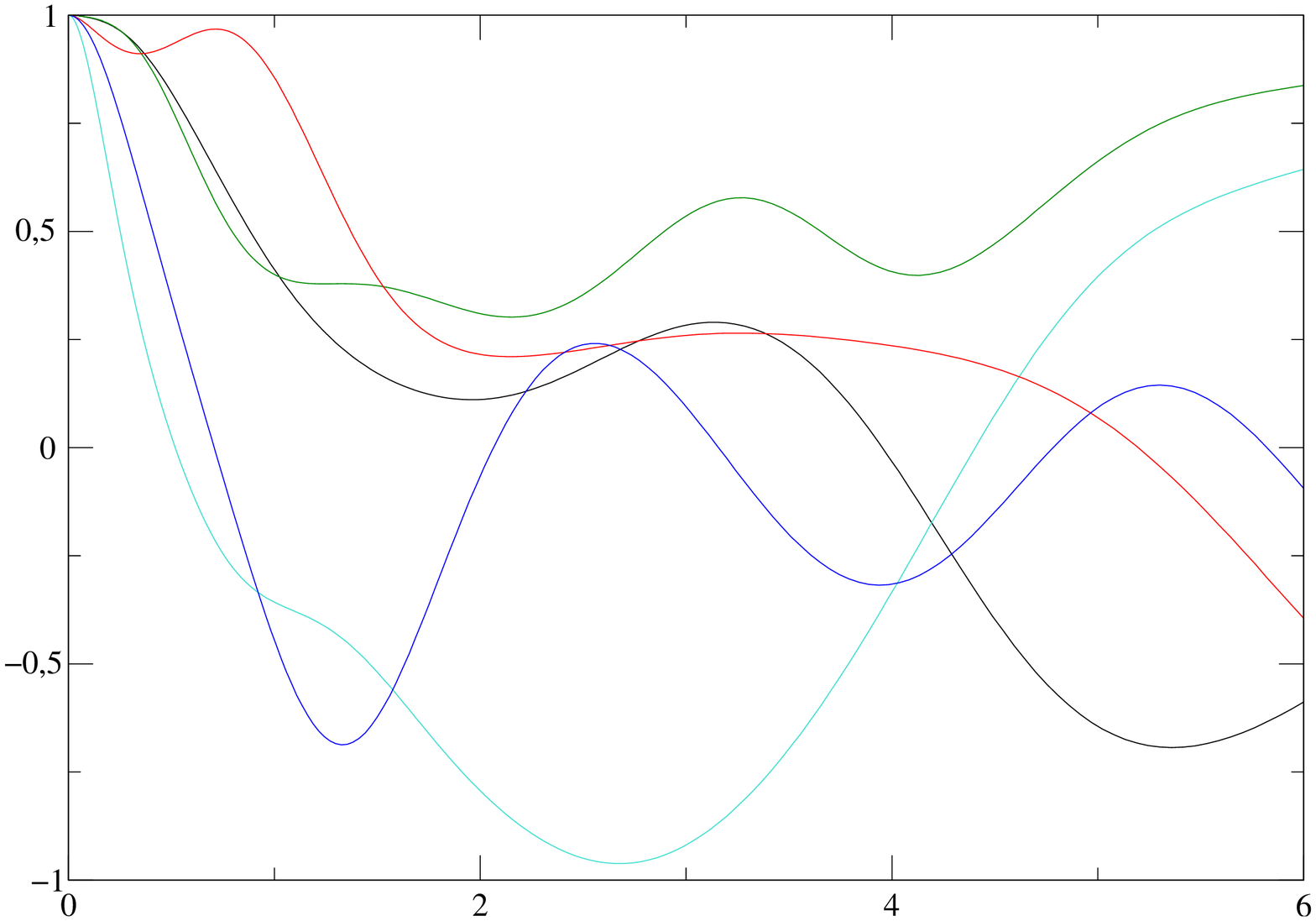, width=0.4\linewidth}\hfill\
\caption{Values of Wilson loops in different planes centered at the origin as a function of the area  of the
 loops (in fm$^2$) for a single configuration in SU(2) lattice gauge theory (left) and for a configuration  of 500 merons
 (right).} \label{brown}  
\end{figure*}

On a lattice, for example, one may realize such a random walk by enlarging
 a  sufficiently large rectangular loop in a fixed configuration by increasing the number of links on each side
 by one.  To the extent that fluctuations produce confinement, the new Wilson loop can be thought of as a small,
 essentially random, step away from the previous loop.  Similarly, for a fixed ensemble of merons or instantons, 
a small increase in the size parameter of the loop, $\rho$, such as the radius of a circular loop, gives a new
 loop whose value can be expected to differ from the previous one by a small, nearly random step.  
Figure \ref{brown} illustrates how the values of differently oriented Wilson loops centered around a common
 point in a single gauge field configuration randomly evolve and become decorrelated as the area increases 
by the order of 1 fm$^2$.   (In the lattice calculation, to damp the ultraviolet fluctuations, we used APE smearing 
\cite{APEC87} with 10 smearing steps, each of which replaced each link by the original link plus 0.5  times the 
sum of staples connecting to that link.)   Each such trajectory as a function of area constitutes a single random
 walk, and an ensemble of such random walk trajectories for an ensemble of configurations then yields the 
distribution of Wilson loops shown in Fig.~\ref{wdlm} for three areas. 

With this motivation, we  now explore the assumption that each trajectory 
is exactly described by a Brownian random walk, where the time parameter, $t$, is an unspecified  increasing
 function of the  size parameter, $\rho$, of the loop. The distribution of an ensemble of such loops, $p(\omega,t)$, 
 is expected to be described by a diffusion equation.
For the case of SU(2) considered in this work, the diffusion
 of the (untraced) Wilson loop occurs on the  group manifold $S^3$.  Due to gauge invariance, the value of the
 (traced) Wilson loop can be identified with  the first polar angle $\vartheta$ on $S^3$. The diffusion process  is 
independent of the two other angles specifying a point on $S^3$.  When applied to functions which only depend 
on  $\vartheta$, the Laplace-Beltrami operator on $S^3$,
\begin{equation} 
 \label{labe0} 
 \Delta_{S^3}= D^{\mu}\partial_{\mu} 
\end{equation}
with the covariant derivative $D_{\mu}$ reduces to 
\begin{equation}
\Delta_{S^3} = \frac{1}{\sin^2 \vartheta}\frac{\partial}{\partial  \vartheta}\sin^2 \vartheta \frac{\partial}{\partial \vartheta}\, . 
\label{labe1}
\end{equation}
The diffusion equation on $S^3$ thus reads
\begin{equation} 
 \label{dieq}
\left(\frac{\partial}{\partial t}-\Delta_{S^3}\right)\,G(\vartheta,t) = \frac{1}{\sin^2 \vartheta}\,\delta(\vartheta)\delta(t)\, .  
\end{equation}
In terms of the eigenfunctions and eigenvalues  of $\Delta_{S^3}$
\begin{equation}
\psi_n =  \sqrt{\frac{2}{\pi}}\,\frac{\sin (n+1)\vartheta }{\sin \vartheta },\quad -\Delta_{S^3}\,\psi_n = n(n+2)\psi_n\, , 
\label{eig}
\end{equation}
the spectral representation 
 \begin{equation}  
\label{spre}
  G(\vartheta,t)= \frac{2}{\pi {\sin \vartheta}}\,\theta(t)\,\sum_{n=1}^{\infty}n\,\sin n\vartheta \,{\rm e}^{-(n^2-1)t}
 \end{equation}  is obtained.

 Up to volume elements, we identify the Wilson loop distribution (\ref{siwi}) with
 the solution of the diffusion equation
\begin{equation} 
 \label{gp}
  p(\cos \vartheta,t) = \sin \vartheta\,\, G(\vartheta,t)\, . 
\end{equation}
This identification does not specify the connection between the size of the loop and the time $t$. However, it  
predicts the shape of the distribution if the expectation value of the Wilson loop and therefore the time $t$ is known. 

\begin{figure*}[t] \hfill\epsfig{file=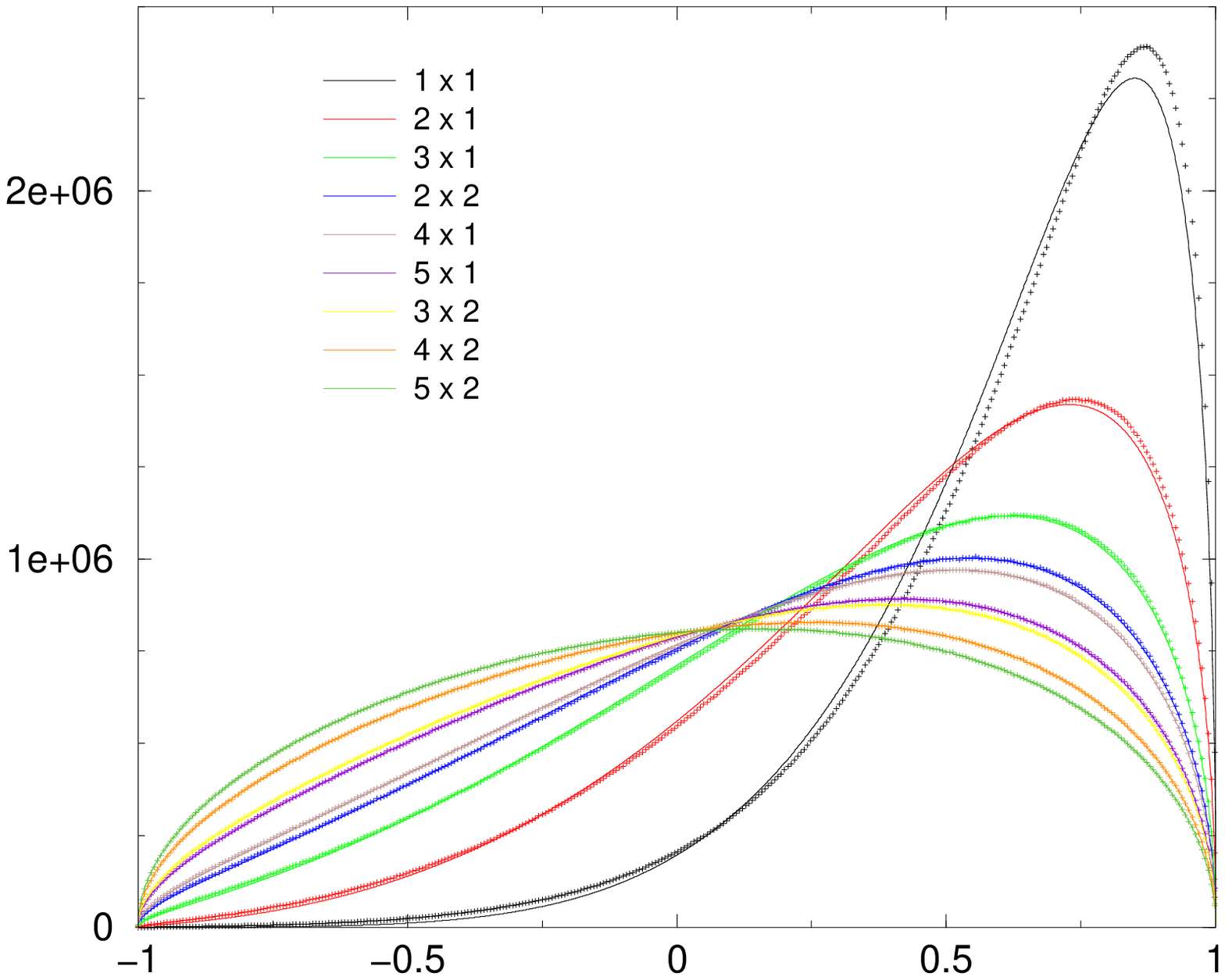, width=.4\linewidth}\hfill\hfill\epsfig{file=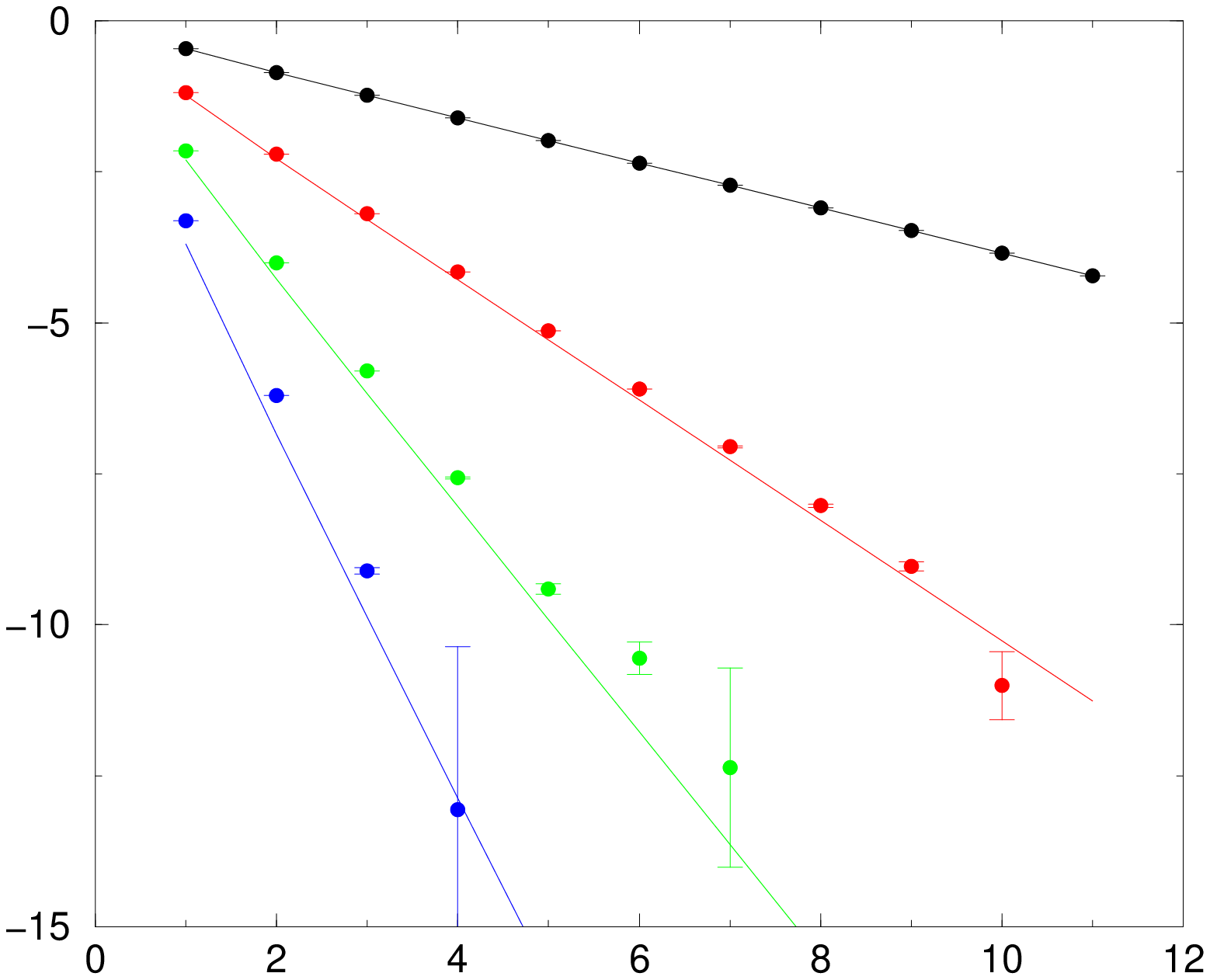, 
width=.4\linewidth}\hfill\
\caption{Approximate Casimir scaling for Wilson loops on a $32^4$ lattice with lattice constant $a=0.13$ fm. 
Left:  Lattice Wilson loop distributions for loops of different aspect ratios with sizes  given in  lattice units. 
The solid lines are   obtained from Eq.~(\ref{spre}) with time $t$ determined by the expectation values.
 Right: Logarithm of Wilson loop as function of the area (in lattice units) for $j=1/2, 1, 3/2, 2$ representations. 
The  straight lines are scaled by the corresponding ratio of the Casimir operators.}
\label{casla}
 \end{figure*}

 As Fig.~\ref{wdlm} and the left panel of Fig.~\ref{casla} demonstrate, the shape of the Wilson loop distribution 
follows  closely the distribution of particles carrying out a Brownian motion on $S^3$ for a wide range of sizes
 and aspect ratios of the loops. In particular, the change in shape from  the peaked distributions for small loops 
 to the equilibrium  distribution, uniform on  $S^3$  for large loops,  $$ t \to \infty,\quad G \to \frac{2}{\pi}$$ is 
correctly predicted, consistent with  Eq.~(\ref{hm}).   \section{Higher representation Wilson loops}  The expectation 
value of an  observable      ${\cal O} (\vartheta)$  for the distribution generated by the diffusion  process on   $S^3$ 
is given by 
\begin{displaymath}
\langle {\cal O} (\vartheta ) \rangle = \frac{2}{\pi} \int ^{\pi}_{0} \sin \vartheta \;{\rm d}\vartheta\; 
 {\cal O} \; (\vartheta) \sum^{\infty}_{n = 1} n \sin n  \;\vartheta \; {\rm e}^{- (n^{2}-1)t}\, .
\end{displaymath} 
Here, we consider the observables associated with the Wilson loops in the  $(2 j + 1)$-dimensional representation of 
 SU(2), which coincide with the eigenfunctions (\ref{eig})
 \begin{eqnarray}
W_{j} (\vartheta) &=& \frac{1}{2 j + 1} {\rm tr} \; \exp\{ 2i \vartheta \left(\begin{array}{ccc}\!\!-j\! & & \\[-0.5em] 
 & \!\ddots\! & \\[-0.5em] & & \!j\! \end{array}\right ) \} = \nonumber\\&=& \frac{\sin (2 j+1)\vartheta}{(2j+1)\,\sin \vartheta} 
=\frac{\sqrt{\pi}}{\sqrt{2}(2 j + 1)} \psi_{2j}(\vartheta)\, .
\label{wj}
\end{eqnarray}
The $j$-dependence of the expectation values 
\begin{equation}  
\label{wex}
\langle W_{j} \rangle = {\rm e}^{- 4j(j+1) t }
\end{equation}
is given in terms of the Casimir operator of SU(2) so that the ratios satisfy Casimir scaling 
\begin{equation}
\label{cassc}
\frac{\ln \langle W_{j_1}\rangle}{\ln \langle W_{j_2}\rangle}=\frac{j_1(j_1+1)}{j_2(j_2+1)} \, . 
\end{equation}
In lattice gauge theory, the validity of Casimir scaling has been demonstrated for SU(2)  \cite{amop84} and  SU(3)  
\cite{bali00} Yang-Mills theory. Approximate  Casimir scaling has also been observed for  ensembles of meron
 configurations \cite{LNT04}. Deviations from Casimir scaling for sufficiently large loops have been observed 
in \cite{defo00, katr02}.  

In addition to the fact that Wilson loop distributions resulting from diffusion on $S^3$ imply Casimir scaling, 
the converse is also true. The Wilson loop operators in the different representations of SU(2) are given by the
 eigenfunctions of $\Delta_{S^3}$. Therefore exact Casimir scaling fixes all moments of the Wilson loop 
distribution up to an overall constant, implying  that the Wilson loop distribution satisfies a diffusion equation 
with the diffusion time being a function of the size of the loop.
This equivalence actually applies more generally 
to  SU($N$). In the spectral representation corresponding to Eq.~(\ref{spre}), the ``character expansion'' of $G$, 
the eigenfunctions corresponding to Eq.~(\ref{eig}) are the characters of the respective representations and
 the eigenvalues are proportional to the corresponding quadratic Casimir operator (cf. \cite{MEON81,ITDR89}).   
 \section{ Screening}
 In the Yang-Mills confining phase,  two regimes in the Wilson loop dynamics can be identified. For small and 
moderate loop size, the Wilson loop distribution  closely  satisfies a diffusion equation and Casimir scaling prevails. 
However, asymptotically, for sufficiently large loop size, complete screening of adjoint (integer) charges and 
screening of the half-integer charges to fundamental charges must occur. Screening of adjoint charges has been
 found in SU(2) lattice gauge theory \cite{defo00,katr02} and for meron ensembles an indication of screening has
 been observed.  The fact that the signal in the screening regime is so weak  suggests that the screening processes 
can be accounted for by a weak perturbation of the diffusion process. Hence, we extend our phenomenological 
 description by adding a drift term to the diffusion equation. To this end we replace the Laplace-Beltrami operator,  
Eq.~(\ref{labe0}), by
 \begin{equation} 
 \label{labem}
\Delta_{S^3}= D^{\mu}\partial_{\mu}\to \tilde{\Delta}_{S^3}= D^{\mu}\partial_{\mu} +
(D^{\mu}\partial_{\mu}V)+ (\partial_{\mu}V) \partial^{\mu}\, , 
\end{equation}
which accounts  for the presence 
of forces  through the potential $V$. As above, due to gauge invariance,
the Wilson loop variable can be identified 
with the first polar angle and the diffusion process must be independent of the two other angles on $S^3$. We 
therefore require$$V=V(\vartheta)\, .$$The solution to the Fokker-Planck equation
\begin{equation} 
 \label{fopl}
\left(\frac{\partial}{\partial t}-\tilde{\Delta}_{S^3}\right)\,\tilde{G}(\vartheta,t) = \frac{1}
{\sin^2 \vartheta}\,\delta(\vartheta)\delta(t) 
\end{equation}
has the spectral decomposition
\begin{equation} 
\label{solfp}
\tilde{G}(\vartheta,t)= \theta(t)\sum_{n=0}^{\infty}\tilde{\psi}_n(\vartheta) \tilde{\psi}_n(0)\, 
{\rm e}^{-\tilde{\lambda}_n t}\,{\rm e}^{V(0)}
\end{equation}
with the eigenfunctions satisfying 
\begin{equation} 
 \label{fpeig} 
 \frac{1}{\sin ^{2}\vartheta} \; \frac{\partial}{\partial \vartheta} \left [ \sin^{2} \vartheta \; \frac{\partial \tilde{\psi}_n}
{\partial \vartheta} +  \sin^{2} \vartheta \; \frac{\partial V}{\partial \vartheta} \, \tilde{\psi}_n \right ] 
=- \tilde{\lambda}_n \tilde{\psi}_n\ .
\end{equation}
The last factor in Eq.~(\ref{solfp}) arises from the nonhermiticity of the differential operator (\ref{labem}). The
 eigenvalue equation possesses the zero mode 
\begin{equation} 
 \label{zm} 
 \tilde{\psi}_{0} = e^{-  V (\vartheta)}\, 
 \end{equation}
and therefore  the asymptotic distribution in the diffusion process is no longer uniform on $S^3$. Rather 
the  equilibrium distribution is determined by the Boltzmann factor.

In order to discuss the properties of the 
screening potential  $V(\vartheta)$, we note that the Wilson loop operators (\ref{wj}) obey the symmetry relation
 for reflection around the equator of $S^3$($\vartheta=\pi/2$)
\begin{equation} 
 \label{wsym}W_{j}(\pi/2-\vartheta)= (-)^{2j}\,W_{j}(\vartheta-\pi/2)\, . 
 \end{equation}
This symmetry property  distinguishes Wilson loops in integer and half-integer representations and must be responsible 
 for their different properties. In order to avoid destroying this distinction, the dynamics of the Wilson loops must
 respect the reflection symmetry, so the potential $V(\vartheta)$  has to be  symmetric
 \begin{equation}
 \label{vsym}  
V(\pi/2-\vartheta)= V(\vartheta-\pi/2)\, .
\end{equation}
Under the influence of $V$, the positive parity ``free'' states $\psi_{2n}$ [cf. Eq.~(\ref{eig})] will mix
 among each other and so will the negative parity states $\psi_{2n+1}$. The symmetry of $V$ prevents
 mixing of even and odd states. The expansion of the Wilson loop operators (\ref{wj}) in terms of the eigenfunctions
 $\tilde{\psi}_{n}$ reads
\begin{equation} 
 \label{wexp}W_{n+\frac{\nu}{2}}({\vartheta})= \sum_{i=0}^{\infty}\omega^{2i+\nu}_{2n+\nu}\,
\tilde{\psi}_{2i+\nu}({\vartheta})\,{\rm e}^{\,\, V(\vartheta)/2}\,, \quad \nu=0,1\,.  
\end{equation}
In general, integer representation Wilson loop operators will contain a non-vanishing admixture from the
 zero mode$$\omega^{0}_{2n}\neq 0\, .$$This  component will dominate at infinite times $t$ [cf. Eq.~(\ref{solfp})], and
 the  expectation values of the Wilson loops in integer representations will become independent of $t$
\begin{equation}  
\label{w2nin}\langle W_{n}(\vartheta) \rangle \to \omega^{0}_{2n}\ .
\end{equation}
The half-integer Wilson loops will also acquire a common asymptotic  behavior, with the  $t$-dependence
 determined by the lowest negative parity eigenvalue [cf. Eq.~(\ref{fpeig})]
\begin{equation} 
 \label{w2nhi}  
\langle W_{n+\frac{1}{2}}(\vartheta) \rangle \to \omega^{1}_{2n+1}e^{-\tilde{\lambda}_1\, t}.
\end{equation}
Thus,  reflection symmetry of $V$ properly  ensures asymptotic screening of the integer Wilson loops
 to adjoint loops  and screening of half-integer loops to the fundamental loops. The transition from Casimir
 scaling to screening is controlled by the strength of $V$. 

We expect the asymptotics of the Wilson loops in
 integer representations to depend on the linear size of the system. For large loops, screening requires the 
formation of two colorless states, so called gluelumps (cf. \cite{defo00}). In turn,  the interaction
 energy [cf. Eq.~(\ref{statq})] of static adjoint charges must satisfy a perimeter law, with the slope determined 
by twice the mass of these objects  $m_{\rm gl}$. This change in the asymptotic behavior requires a time 
dependent  potential  \begin{equation}  \label{td}V=V(\vartheta,t)\end{equation}which  asymptotically decreases
 with time. For large $t$,  from the area law for half-integer loops  $$t\sim \rho^2\, .$$ The correct asymptotics 
for integer loops is reached, provided 
\begin{equation}  
\label{alas}
V(\vartheta,t) \to V_{\infty}(\vartheta)\,e^{-\mu\sqrt{t}}\,, \quad \mu\sim 2 m_{\rm gl}\,.
\end{equation}
When dynamical fermions are present also  half-integer Wilson loops will satisfy a perimeter rather than an area law.
 In this framework, this essential change due to fermions can be incorporated by admitting potentials $V$ which 
violate the symmetry relation, Eq.  (\ref{vsym}). In this case, one expects all the Wilson loops to approach a common 
asymptotic  behavior determined by the mass of the lightest meson. For heavy quarks and circular loops the potential
 can be calculated \cite{bzlt}, with the result 
\begin{equation}  
\label{qupo}
V_{\rm q}(\vartheta,t) \sim (\vartheta -\pi/ 2)\, t^{-7/4}\, {\rm e}^{-\mu \sqrt{t}}\,,
\quad \mu\sim 2m_{\rm q}\,, 
\end{equation}with the quark mass $m_{\rm q}$. The quark induced potential $V_{\rm q}$ is antisymmetric
 around the equator of $S^3$ and therefore turning on $V_{\rm q}$ mixes even states ($\tilde{\psi}_{2n}$)  and
 odd states ($\tilde{\psi}_{2n+1}$)\, . 
 \section{ Wilson Loop Effective Action}
Our results suggest treating the Wilson
  loop variables (\ref{wilo1})
\begin{equation} 
 \label{athe} 
 W = \cos a
\end{equation} 
as quantum mechanical
 degrees of freedom with the following effective action
 \begin{equation}
S_{\rm eff}[a]=\int {\rm d}t \Big\{\frac{1}{4}\dot{a}^2 + U(a)\Big\}\, .   
\label{act}
\end{equation}
For a quantum mechanical system the path integral
 \begin{eqnarray}
  g(t^{\prime},t;\vartheta^{\prime},\vartheta) &=& \int_{a(t^{\prime})=\vartheta^{\prime}}^{a(t)=
\vartheta}{\rm d}[a]\;{\rm e}^{-S[a]}\,, \nonumber\\
{\rm d}[a] &=& \prod_i \int_0^{\pi}\sin^2a_i\: {\rm d}a_i
\label{path1}
\end{eqnarray}
satisfies the diffusion equation on $S^3$
\begin{equation} 
 \label{dieqw}
\left(\frac{\partial}{\partial t}-\Delta_{S^3}+U\right)\,  g(t^{\prime},t;\vartheta^{\prime},\vartheta) = 
\frac{1}{\sin^2 \vartheta}\,\delta(\vartheta-\vartheta^{\prime})\,\delta(t-t^{\prime})\, . 
\end{equation}
In terms of $g$, the Wilson loop distribution is expressed as
\begin{eqnarray} 
 p(\vartheta,t) &=& \frac{ g(0,t;0,\vartheta)\;\int {\rm d}\vartheta_{\infty}\,\sin^2\vartheta_{\infty}\; 
g(t,\infty;\vartheta,\vartheta_{\infty})}{\int {\rm d}\vartheta_{\infty}\,\sin^2\vartheta_{\infty}\; g(0,\infty;0,\vartheta_{\infty})} =
 \nonumber\\
&=& g(0,t;0,\vartheta)\frac{\chi_{0}(\vartheta)}{\chi_{0}(0)}\, .  
\label{gg}
\end{eqnarray}
In the evaluation of the integral we have used the spectral representation of the Green function $g$. At infinite
 times, only the ground state of the system
 \begin{equation}  
\label{gst}
 (-\Delta_{S^3}+U) \chi_{0}(\vartheta)=0 
\end{equation}
contributes. It is straightforward to verify that the Wilson loop distribution satisfies the Fokker-Planck
 equation
\begin{equation} 
 \label{foplp}
\left(\frac{\partial}{\partial t}-\tilde{\Delta}_{S^3}\right)\, p(\vartheta,t) = \frac{1}{\sin^2 \vartheta}\,\delta(\vartheta)\delta(t)\, ,  
\end{equation}
with the potential $V$ appearing in $-\tilde{\Delta}_{S^3}$ [cf. (\ref{labem})] given by
 \begin{equation} 
 \label{pv}
  \, V(\vartheta) = -\ln \chi_{0}^{2}(\vartheta)\, .
\end{equation}
The  potential $\, V$, which may in principle be obtained  from a fit to the Wilson loop distribution, 
determines the potential $U$ appearing in the Wilson loop effective action
\begin{equation} 
 \label{vw}  
U(a) = \, \frac{{\rm d}V}{{\rm d}a}\Big[-\cot a +\frac{1}{4}  \, \frac{{\rm d}V}{{\rm d}a}\,\Big]-\frac{1}{2} \, 
\frac{{\rm d}^{2}V}{{\rm d}a^{2}} \, .
\end{equation}
Reflection symmetry of  $\, V$ around $a = \pi/2$ implies that of $U$ and vice versa, provided the
 ground state wavefunction $\chi_0$ is also even under reflections.  The results of our investigations
 can be described by the following effective action of the Wilson loop variables depending on the size
 parameter $\rho$
\begin{equation}
S_{\rm eff}[a]=\int {\rm d} \rho\: \tau^{-1}(\rho)\, \left\{\frac{1}{4}\Big(\tau (\rho)\, \frac{{\rm d}a}{{\rm d}\rho}
\Big)^2+ U(a)\right\}\, .   
\label{act2}
\end{equation}
The  function
\begin{equation} 
 \label{tau}
\tau(\rho) =\Big(\frac{{\rm d}t}{{\rm d}\rho}\Big)^{-1} 
 \end{equation} 
remains unspecified.  Irrespective of the choice of $\tau$, the Wilson loop distributions generated by
 this action  are solutions of the Fokker-Planck equation (\ref{foplp}). In the limit $U=0$, Wilson loops exhibit
 Casimir scaling. In this limit, the action possesses a shift symmetry $a(\rho) \to a(\rho)+\delta a $, 
corresponding to translational symmetry on $S^3$. Clearly, any $a$-dependent $V$ destroys this symmetry
 and thereby gives rise to violations of Casimir scaling.

The effective action separates the dynamics of diffusion, 
 with the associated phenomena of Casimir scaling and screening,  from the  Wilson loop dynamics governing
 the size dependence, with the appearance of perimeter or area laws incorporated in  $\tau(\rho)$.
\section{  Connection to Yang-Mills Theory  }
In  the following discussion, we will indicate the connection of our phenomenological findings to SU(2) Yang-Mills
 theory. We will restrict our considerations to circular loops (radius $\rho$) and discuss first the limit of small loops. 

 In the short-time limit of vanishing Wilson loop size, a regular potential $U$  becomes irrelevant. The Wilson loop
 distribution, Eq.~(\ref{spre}),
\begin{equation}  
\label{smti}
p(\vartheta,t)\approx  \frac{2}{\pi \vartheta}\int^{\infty}_{0} {\rm d} n \: n\, \sin n \vartheta \;{\rm e}^{-  n^{2}t} = 
\frac{1}{2\sqrt{\pi}\, t^{3/2}} \; {\rm e}^{- \vartheta^{2}/4 t}
\end{equation}
is obtained by  diffusion in $I\!\!R^3$, the tangent space of SU(2).
For small sizes, the Wilson loops can
 be calculated perturbatively. To leading order, the SU(2) Wilson loops coincide with the free U(1)$^3$ 
gauge theory. In U(1) gauge theory the Wilson loop distribution  is given by($\vartheta\in I\!\!R$)
\begin{eqnarray}  
p_{\rm U(1)}(\vartheta,\rho)&=&\frac{1}{Z}\int {\rm d}[A]\;\; \delta(\vartheta - g\oint_{\cal C} d x_\mu A_\mu )
\;{\rm e}^{-S[A]} = \nonumber\\
&=& \frac{1}{2\pi\sqrt{\kappa}}\exp\left(-\frac{\vartheta^2}{4\pi \kappa}\right)  
\label{u1}
\end{eqnarray}
with
\begin{equation}
\kappa=\frac{g^2}{2\pi}\oint_{\cal C} d x_\mu\oint_{\cal C} d y_\nu\, D_{\mu\nu}(x-y)\, .
\label{kap}
\end{equation}
The singularity in the Feynman propagator,   $$D_{\mu\nu}(x)=\frac{1}{4\pi^2 x^2}\,\delta_{\mu\nu}$$
can be regularized by Gaussian smearing of the point charge  over distances $\sim \epsilon$ perpendicular
 to the loop plane. To leading order in $\epsilon$
 \begin{equation}
\label{fcirc}
\kappa = \frac{g^2}{4\sqrt{2\pi}}\frac{\rho}{\varepsilon}\quad.
\end{equation}
In SU(2) Yang-Mills theory, for small loop size, the Wilson loop distribution,  given by 
\begin{equation} 
 \label{smsi}  
p(\vartheta,\rho)\,\vartheta^2\, {\rm d}\vartheta = \prod_{i=1}^{3}\, p_{\rm U(1)}(\vartheta_i,\rho)\, {\rm d}\vartheta_i\, ,
\end{equation}
is the distribution generated in a diffusion process in $I\!\!R^3$. From comparison with (\ref{smti}), we 
conclude 
\begin{equation}  
\label{ter} 
 \rho \to 0\,,\quad t=\frac{g^2\sqrt{\pi}}{4\sqrt{2}\,\epsilon}\,\rho \, ,
\end{equation}
and therefore, for small Wilson loops  the effective action is given by
\begin{equation}
 \label{ssm}
S_{\rm eff}[a]= \frac{\sqrt{2}\, \epsilon}{\sqrt{\pi}g^2}\int \, {\rm d} \rho\, \Big( \frac{{\rm d}a}{{\rm d}\rho}\Big)^2 \, . 
\end{equation}
In the limit of large loop sizes, the fundamental loops obey an area law which, in the Casimir scaling
 regime [Eq.~(\ref{spre})] requires the identification 
\begin{equation} 
 \label{arla} 
\rho \to \infty\, , \quad  t = \frac{\sigma \pi}{3} \rho^2\, .
 \end{equation}
Thus for large loop sizes
\begin{equation}  
\label{laef}
S_{\rm eff}[a]=\frac{3}{8\pi \sigma} \int \frac{{\rm d}\rho}{\rho} \Big(\frac{{\rm d}a}{{\rm d}\rho}\Big)^2 \, .
\end{equation}
Assuming the logarithm of the Wilson loops to be a sum of  constant, perimeter and area terms, the function
 $\tau(\rho)$ respecting the two limits Eq.~(\ref{ter}), Eq.~(\ref{arla}) is given by 
\begin{equation} 
 \label{taufit} 
 \tau(\rho)=\left[\frac{g^2\sqrt{\pi}}{4\sqrt{2}\,\epsilon} +\frac{2\pi\sigma}{3}\rho\right]^{-1}\, .
\end{equation}
Finally, we  compare the effective action with the Yang-Mills action. To this end,  we choose the gauge in which
  circular loops appear as elementary degrees of freedom, and assume the loops  are located in the ($x,y$)-plane
 and centered around the origin. We choose cylindrical coordinates $\rho,\varphi,z,t$ and transform to the
 azimuthal gauge \cite{bzlt} with the $\varphi$-component of the gauge field
\begin{equation} 
 \label{aziga}
 A_{\varphi}(x) = \frac{1}{g\pi \rho}\, a\left(\rho,z,t\right)\frac{ \tau_{3}}{2}
 \end{equation}
being diagonal and independent of  $\varphi$. With this definition, the gauge field $a$ is the Wilson
 loop [cf. Eqs.~(\ref{wilo1}), (\ref{athe})].  In field strength components $F^3_{\varphi \alpha}$ containing
 derivatives of $A_\varphi$, no commutator terms are present, e.g.
\begin{equation}  
\label{frp} 
 F_{\rho\varphi}^3=\frac{1}{\rho}\left(\frac{1}{g\pi}\partial_{\rho} a - \partial_{\varphi}A_{\rho}^3\right)\,.
 \end{equation}
The effective action of the Wilson loop variables is defined by
\begin{equation} 
 \label{eaym}  {\rm e}^{-S_{\rm eff}^{\rm YM}[a]}=\int {\rm d}[A^{\prime},a^{\prime}] {\rm e}^{-S_{\rm YM}
[A^{\prime},a^{\prime}]}\prod_{\rho}\delta\Big(a^{\prime}(\rho,0,0)-a(\rho)\Big)\, .
\end{equation}
In the Yang-Mills action
\begin{eqnarray} 
 \label{SYM}  
S_{\rm YM} &=& \int \rho\, {\rm d}\rho\, {\rm d}\varphi\, {\rm d}z\, {\rm d}t\,
 \\&& \times\Big\{ \frac{1}{2\pi^2 g^2\,\rho^2}
\Big[(\partial_{\rho} a )^2 +(\partial_{z} a )^2 +(\partial_{t} a )^2\Big] + {\cal L}^{\prime}\Big\}\,, \nonumber
 \end{eqnarray}
due to the $\varphi$ independence of $a$, no bilinear mixing of the Wilson loop variables with the other
 fields occurs. ${\cal L}^{\prime}$  contains the interaction of the Wilson loop variables with the other fields. 
 The integration over the Wilson loop variables in Eq.~(\ref{eaym}) includes, as in Eq.  (\ref{path1}), the Haar
 measure. For comparison with the results in Eq.~(\ref{ssm}) and Eq.~(\ref{laef}) we observe that the
 effective action of the Maxwell theory is obtained by dropping  the Haar measure in Eq.~(\ref{eaym}), 
integrating $a^{\prime}$ over the real axis, and disregarding the $a^{\prime}$-independent term 
${\cal L}^{\prime}$ in Eq.~(\ref{SYM}). Integrating out the Wilson loop variables at $(z,t)\neq (0,0)$ apparently
 modifies the volume element in the effective action. Propagation of the Wilson loop variables in  Maxwell 
theory occurs in (3+1)-dimensional space-time and therefore actively involves the degrees of freedom 
 located at $(z,t)\neq (0,0)$. As Eq.~(\ref{laef}) suggests, this is not the case for large Wilson loops
 in Yang-Mills theory. Here the integration over the degrees of freedom other than $a(\rho,0,0)$ renormalizes
 the bare action but does not change the volume element. The nearest neighbor interaction seems to be
 operative only for small loops.  This can be interpreted as a consequence of the formation of  a flux tube,
 which turns the effective action into  that of a (1+1)-dimensional system. Indeed, evaluation of $\kappa$, 
Eq.~(\ref{kap}), using the two-dimensional propagator $D_{\mu\nu}(x)\sim \delta_{\mu\nu} \ln x^2 $ yields the 
effective action, Eq.~(\ref{laef}). 

Our last topic concerns the evaluation of the Wilson loop distribution 
[Eq.~(\ref{siwi})] in SU(2) Yang-Mills lattice gauge theory.   In the strong coupling expansion the distribution
 of the values of a square   Wilson loop of area ${\cal A}$  is given by
\begin{eqnarray} 
 \label{ stc}
  p(\cos \vartheta,t) &=& \frac{2}{\pi}\sin \vartheta \,\left[1+4 \cos \vartheta \, e^{-t \ln g^2} + \right. \\
&& \qquad\left.+ ( 4\cos^2 \vartheta -1)\,e^{-16(\sqrt{t}-1)\ln g^2} + \ldots\right]\nonumber
\end{eqnarray}
where, in terms of the lattice constant $a$,
 \begin{equation} 
 \label{tt}
t={\cal A}/a^2.
\end{equation}
We have kept only the leading contributions in $1/g^2$ to the  area and perimeter dependent terms.
  To this order,  the distribution function gives rise to non-vanishing expectation values for loops in the
 fundamental and the adjoint representation. The area term arises from tiling of the minimal surface 
enclosed by the loop, the perimeter term from tiling the minimal surface of a rectangular ``tube"  of 
volume $\sim 4 \sqrt{\cal A}\, a^2$ along the loop.  As expected, in the distribution  the area and perimeter
 contributions are  distinguished by the reflection symmetry in $\omega=\cos \vartheta$. Symmetric
 and antisymmetric components $p(\omega,t)\pm p(-\omega,t)$ contain only terms which satisfy a perimeter
 and an area law respectively. For large loops therefore, $p(\omega,t)$ becomes an even function in $\omega$.
 We also observe that, to leading order, strong coupling implies  screening of Wilson loops in integer
 representations. Thus in strongly coupled lattice gauge theory, Casimir scaling is not a natural limit. Since 
Casimir scaling implies an area law for loops in every representation, tiling by including contributions from
 links transverse to the loop must be  suppressed by dynamics beyond the strong coupling limit. We thus
 reach in lattice gauge theory the same conclusion as in the continuum theory discussed above. In the
 limit of Casimir scaling, the effective theory has to be two-dimensional.  Following the line of arguments
 given above,  we can reach  this limit formally by assuming  that links transverse to the loop (in the 1-2 plane)
 are not affected by the presence of the loop.   The distribution function [Eq.~({\ref{siwi})]  is thereby reduced 
to a  functional integral over the links in the plane of the loop $U_{12}$ and planes parallel to it  
\begin{equation}  
\label{tdc} 
p(\cos \vartheta ,t)\approx \frac{\int {\rm d}[U_{12}]\,e^{-S_{12}[U_{12}]}\,\delta\left(
 \cos \vartheta-\frac{1}{2} {\rm tr}\prod_{\alpha} U_{12}^{\alpha}\right)}{\int {\rm d}[U_{12}]\,e^{-S_{12}[U_{12}]}}\, . 
\end{equation}
To simplify the action
 \begin{equation} 
 \label{s12}
S_{12}[U_{12}]= -\frac{2}{g^2}\,\mbox{tr}\sum_{n_1,n_2,n_3,n_4}P_{12}(n_1,n_2,n_3,n_4)\, ,
\end{equation}
we assume that  the loop affects the plaquettes $P_{12}$ only in 1-2 planes within a distance $\lambda $
 from the plane of the loop and that the contributions to  the action from all of these planes are the same. 
With these approximations one may construct a crude but suggestive model of the flux tube picture of 
confinement. With 
\begin{equation}  
\label{s12a}
S_{12}[U_{12}]\approx -\frac{2}{g_{(2)}^2 a^2}\,\mbox{tr}\sum_{n_1,n_2}P_{12}(n_1,n_2)\, ,
\end{equation}
we arrive at  two-dimensional Yang-Mills theory with coupling constant
\begin{equation} 
 \label{cc}
g_{(2)}^2 = \frac{g^2}{\pi\lambda^2} .  
\end{equation}
In two-dimensional  Yang-Mills theory, the partition function and various observables can be evaluated
 in closed form, cf. \cite{Migdal75,GRWI80} and some recent applications thereof \cite{Tierz04}. We obtain
 the following expression for the Wilson loop distribution
\begin{equation} 
 \label{wd2} 
 p(\cos \vartheta ,t)\approx \frac{2}{\pi}\sum_{n=1}^{\infty}n\, \sin n\vartheta \left( \frac{I_n(2\beta)}{I_1(2\beta)}\right)^t , 
\end{equation}
in terms of the modified Bessel functions $I_n$ depending on the two-dimensional (inverse) coupling constant 
\begin{equation} 
 \label{bet}  
\beta = \frac{ 2\pi \lambda^2}{g^2 a^2}
 \end{equation}
and $t$ as given in Eq.~(\ref{tt}). In the continuum limit
\begin{equation} 
 \label{coli} 
\beta \rightarrow \infty\, ,\quad \frac{I_n(2\beta)}{I_1(2\beta)}\rightarrow 1-\frac{n^2-1}{4\beta}\,   
\end{equation}
we finally obtain
 \begin{equation}  
\label{wd2c}
  p(\cos \vartheta ,t)\approx \frac{2}{\pi}\sum_{n=1}^{\infty}n\, \sin n\vartheta e^{-(n^2-1)\sigma {\cal A}/3} ,
 \end{equation}
with the string constant in the fundamental representation
\begin{equation}  
\label{sig} 
 \sigma = \frac{3 g^2}{ 8\pi\lambda^2}\,.
\end{equation}
In this way, we reproduce  the distribution  obtained from diffusion [cf. Eq.~(\ref{spre})] and thereby
 confirm the intimate connection between Casimir scaling and two-dimensional dynamics. Whereas
  confinement in  two dimensions is of kinematical origin,  Casimir scaling is obtained  only in the continuum
 limit.  In the strong coupling limit ($\beta \to 0$), the $n^2-1$ dependence on the degree of the
 representation in Eq.~(\ref{wd2c}) is replaced by a linear $n$ dependence.  
 \section{Conclusion}
In this work we have presented a phenomenological analysis of the distribution of Wilson loops in SU(2) 
Yang-Mills theory. We have shown that Wilson loop distributions calculated in lattice gauge theories or 
obtained in model studies can be succinctly represented by distributions resulting from diffusion of Wilson
 loops on the group manifold. This description is applicable to small loops  in  the regime of asymptotic
 freedom  as well as  to large loops in the confinement regime.  As preliminary results \cite{bzlt} indicate,
 it applies equally well in the deconfined phase. 

Casimir scaling of Wilson loops in different representations
 is a natural limit in this description. It is an exact property of Wilson loops if and only if  the diffusion occurs
 in the absence of external forces. It reflects the translational symmetry on the group manifold. This 
equivalence of Casimir scaling and free diffusion remains true for SU($N$). 

Screening processes can
 be accounted for by a potential term in the diffusion equation,  or equivalently, a drift term in the 
 Fokker-Planck equation. The crucial distinction in the screening process between fundamental and
 adjoint loops is formulated in this framework as a symmetry property of the potential term. A 
reflection symmetry protects the area law of large fundamental loops from being violated 
by screening  processes. The initial condition in the diffusion process breaks this symmetry by 
singling out the point on the group manifold where the source is located. Consequently, the  
solution of the diffusion equation makes this symmetry manifest only at asymptotic times, that is, 
 for asymptotically large loops. This analysis suggests the existence of  a similar symmetry to 
distinguish fundamental and adjoint loops in the underlying gauge theory.  The associated 
symmetry transformations  must include  a $Z_2$ center reflection of Wilson loops. 
Unlike the  $Z_2$ symmetry related to Polyakov loops, this center symmetry exists  irrespective 
of a  compact direction in space-time.

The formulation of our results in  terms of an effective Wilson 
loop action separates confinement and screening dynamics. Confinement is primarily connected to
 the properties of the kinetic term in the action. The existence of an area law is related to the 
effective two-dimensional structure of the kinetic term for large loop sizes. This remains true 
as long as the potential respects a reflection symmetry.

The strength of the potential term 
controls the deviations from Casimir scaling. For a non-vanishing potential term, Wilson loops
 in half-integer representations approach a common asymptotic behavior and so do Wilson 
loops in integer representations. If the potential violates the reflection symmetry,  as induced
 for example by dynamical fermions, an area law  can only show up, if at all, at intermediate sizes. 
Asymptotically, the distinction between half-integer and integer representations in the size dependence
 of the loops  disappears. 

Extension of this work to SU(3) and more generally to SU($N$) is of interest. In
 SU($N$), the starting point will be a diffusion equation for $N-1$ ``radial'' variables with associated 
Jacobians (cf. \cite{MEON81,LNT94}). These variables correspond to the $N-1$ Wilson loops that are 
elements of the Cartan subalgebra. The symmetry of the potential is crucial for the distinction of 
confined and nonconfined variables. For SU(3), in a different context, the choice of variables and 
the symmetry analysis has been carried out \cite{LST95}. It will be of interest to see whether the 
suppression of screening processes in SU(3) as compared to SU(2) can be related to differences 
in the center symmetry. For $ N\geq 4$, a new element in the dynamics appears. The string tension 
of strings of different $Z_N$ charge may be chosen independently of each other. As a consequence, 
Casimir scaling might be violated or new scaling laws might emerge as string or supersymmetric
 theories suggest \cite{ARSH03} and lattice calculations indicate \cite{DPRV02}. A phenomenological 
symmetry analysis of the potential term in the corresponding Fokker-Planck equation may also offer
 a new perspective on the dynamics of Wilson loops and help to clarify  their large-$N$ limit.  
\vskip .5cm 
This work was supported in part by funds provided by the U.S. Department of Energy (D.O.E.) under
 cooperative research agreement DE-FC02-94ER40818.    
 \bibliographystyle{unsrt}
 
 \end{document}